\documentclass[9pt,twocolumn,twoside]{revtex4-1}

\usepackage{graphicx}% Include figure files
\usepackage{subfigure}% Include subfigure files
\usepackage{dcolumn}% Align table columns on decimal point
\usepackage{bm}% bold math \usepackage{mathrsfs}
\usepackage{subeqnarray}
\usepackage{amsmath}
\usepackage{color}
\usepackage{esvect}
\usepackage{ulem}

\newcommand{\LN}{LiNbO$_3$ }
\newcommand{\Fig}[1]{Fig.~\ref{#1}}

\emergencystretch=\maxdimen
\begin{document}

\title{Counter-propagating photon pair generation \\ in a nonlinear waveguide}

\author{Kai-Hong Luo$^{\star}$, Vahid Ansari, Marcello Massaro, Matteo Santandrea,  Christof Eigner,  \\
Raimund Ricken, Harald Herrmann, and Christine Silberhorn}

\affiliation{Department of Physics and CeOPP, University of Paderborn,Warburger Strasse~100, 33098 Paderborn, Germany}

\affiliation{$^{\star}$E-mail:khluo@mail.uni-paderborn.de.}

\begin{abstract}
Counter-propagating parametric conversion processes in non-linear bulk crystals have been shown to feature unique properties for efficient narrowband frequency conversion. In quantum optics, the generation of photon pairs with a counter-propagating parametric down-conversion process (PDC) in a waveguide, where signal and idler photons propagate in opposite directions, offers unique material-independent engineering capabilities. However, realizing counter-propagating PDC necessitates quasi-phase-matching (QPM) with extremely short poling periods. Here, we report on the generation of counter-propagating single-photon pairs in a self-made periodically poled lithium niobate waveguide with a poling period on the same order of magnitude as the generated wavelength. The single photons of the biphoton state bridge GHz and THz bandwidths with a separable joint temporal-spectral behavior. Furthermore, they allow the direct observation of the temporal envelope of heralded single photons  with state-of-the art photon counters.
\end{abstract}

\maketitle

%%%%%%%%%%%%%%%%%%%%%%%%%%  body  %%%%%%%%%%%%%%%%%%%%%%%%%%
\section{Introduction}
Nonlinear optical processes exploiting quasi-phase-matching (QPM) \cite{HumQPM2007, ArieLPR2010} in periodically poled $\chi^{(2)}$ materials provide a versatile means for numerous frequency conversion devices, e.g. optical parametric oscillators (OPOs)  \cite{GiordmainePRL1965}. Tremendous progress has been achieved by exploiting specifically tailored poling patterns. Among them, domain patterns with ultra-short poling periods have come into the focus of scientific attention as these can be used for counter-propagating processes with applications in the classical nonlinear \cite{KangOL1997,GuJOSAB1998} as well as in the quantum domain \cite{BoothPRA2002, URenLP2005, RavaroJAP2005, LancoPRL2006, JanPRA2008, SuharaIEEE2010, GongPRA2011, BoucherPRA2014, GattiPRA2015,GattiPRA2018, BashkanskySPIE2016,LatypovQE2017,SaraviPRL2017} . For example, efficient mirrorless OPOs \cite{HarrisAPL1966} in bulk KTP crystals with ultra-short periods have been demonstrated \cite{CanaliasNP2007,ZukauskaSR2017, ViottiCleo2019}. However, different from such classical devices, harnessing counter-propagation for quantum applications is still in its infancy. To date, detailed experimental studies of counter-propagating parametric down-conversion (PDC) and the quantum properties of the generated photon pairs have not yet been performed. Neither efficient counter-propagating separable quantum state generation nor a detailed study of temporal and spectral properties of such two-photon states have been demonstrated which is mostly due to practical limitations due to low efficiency. To overcome these limitations waveguide devices promise many benefits to achieve the required efficiencies by exploiting the strong confinement of the fields over a long interaction length.

In a counter-propagating PDC source, one of the generated photons travels in the forward direction along with the pump beam while the other is back propagating against the pump direction. QPM for such a process requires a poling pattern with ultrashort periods. It was theoretically proven that such a process can generate separable quantum states with different bandwidths \cite{URenLP2005}. The spectral and temporal properties of generated photons are independent from each other so that the detection of one photon yields no information whatsoever about the other one. Such separable twin-photon states \cite{CaspaniLSA2017,AnsariOp2018} lie at the heart of many secure quantum computation and networks. For example, to link stationary and flying qubits in quantum networks \cite{NorthupNP2014, RutzPRAP2017}, separable quantum states, i.e., spectrally decorrelated photon-pair sources, are required which provide one photon with narrow bandwidth to address an atomic quantum memory (typically in the ultraviolet, visible or near-infrared range) and the second one having broader bandwidth to propagate as short wave-packet through the optical fiber network at telecom wavelength.  However, the major obstacle for an efficient generation of such separable states via counter-propagating PDC is the technological challenge related with the fabrication of ultrashort poling periods.

The most efficient co-propagating PDC sources use periodically poled waveguides. An efficiency improvement by about 4 orders of magnitude compared to the efficiency obtainable in bulk crystal can be obtained \cite{TanzilliEP2002,ZielnickiJMO2018}. In recent years, the generation of counter-propagating decorrelated qubits in integrated waveguides has been considered \cite{MosleyPRL2009, ChristOE2009}, but only a few experimental investigations have been carried out. The main reason is directly associated with periodic poling technology: a periodic poling with ultra-short periods in the sub-$\mu$m range, which reaches resolution limit of optical lithography. Moreover, many fundamental aspects of this vacuum seeded counter-propagating photon pair generation and an intuitive interpretation of this process, which is entirely different from the generation of narrowband single photons inside a cavity \cite{FeketePRL2013, LuoNJP2015}, are still to be clarified although in recent years several theoretical models \cite{GattiPRA2018} have been developed.

In this paper, our results on counter-propagating photon pair generation in Ti-diffused LiNbO$_3$ waveguides are presented. We demonstrate that  with on optimization of the underlying technologies periodically poled waveguides with periods of 1.7 $\mu$m can be fabricated. With these devices counter-propagating photon pair generation exploiting 5-th order QPM could be demonstrated and a detailed quantum spectral and temporal characterization. Apart from gaining inside into the fundamental process of counter-propagating pair generation, the devices with the ultra-short poling periods also complement the toolbox of lithium niobate components \cite{SharapovaNJP2017} for future advanced quantum circuits, and the substantially improved poling technology paves the way towards integrated quantum photonic circuits based on lithium niobate thin films \cite{BoesLPR2018}.

\section{Counter-propagating QPM}

%%%%%%%%%%%%%%%%%%%%%%%%%%%%%%%%
\begin{figure*}[htb]
\center
\includegraphics[width=0.85\textwidth]{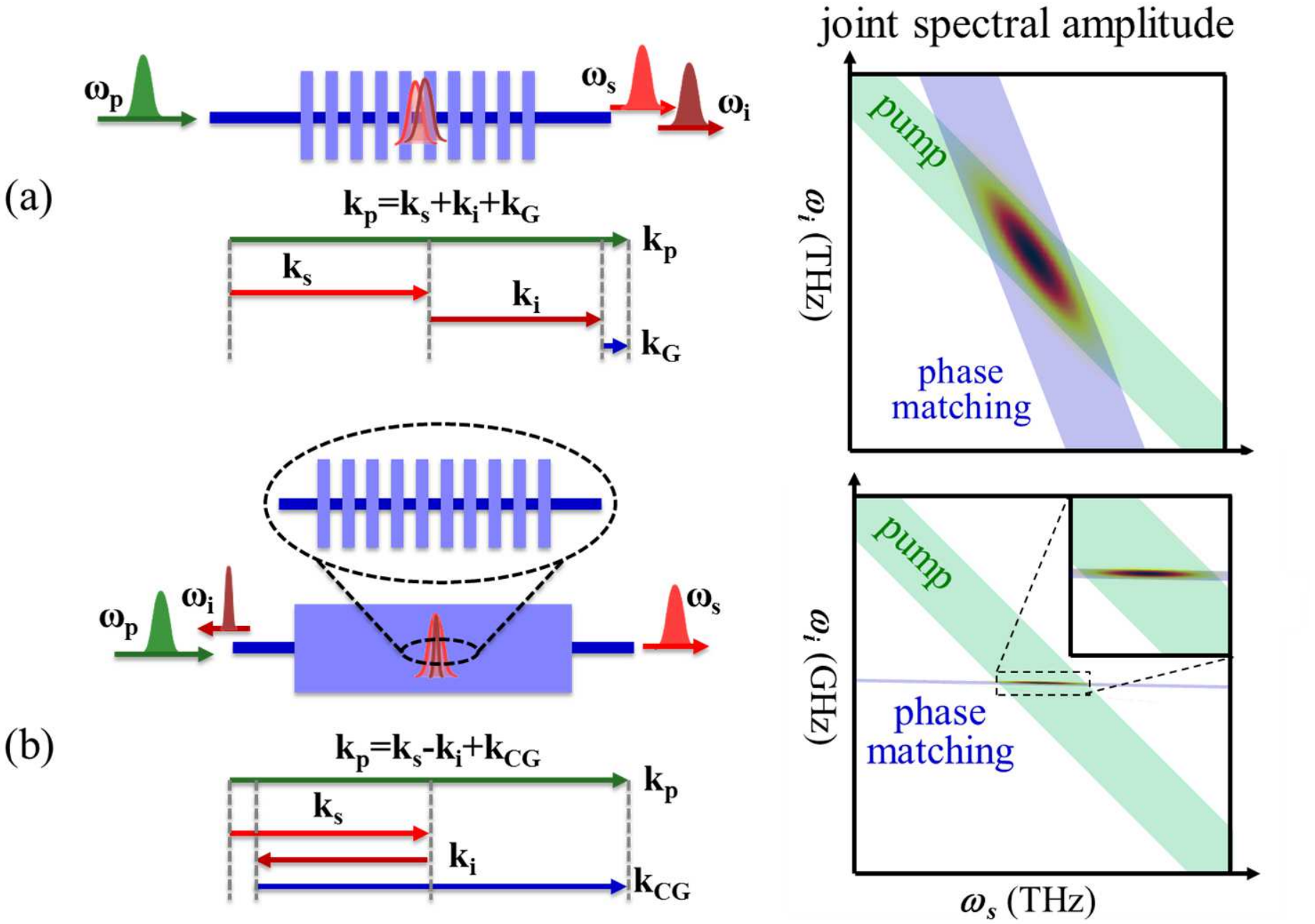}
\caption{\label{QPM_JSI}
Sketches of photon pair generation, vector diagrams of quasi-phase-matching (QPM) interactions, and joint spectral amplitudes (JSA) of various PDC processes in a periodically poled ferroelectric waveguide. (a) Conventional PDC via co-propagating phase matching ($k_p =k_s+k_i+k_G$), where the signal and idler photons co-propagate along the same direction as the pump. The JSA is determined by the pump distribution and the phase-matching dispersion inside the medium. Because of the dispersion properties of the nonlinear material, the JSA still shows  spectral correlations between signal and idler. (b) Counter-propagating phase-matching ($k_p =k_s-k_i+k_{CG}$) in a waveguide with a much shorter poling period, where the generated idler photon is counter-propagating to the pump photon. Because of the narrow and almost horizontal phase-matching function in the JSA diagram, the PDC process generates decorrelated bi-photon states with a narrow spectral width for the idler photon.
}
\end{figure*}
%%%%%%%%%%%%%%%%%%%%%%%%%%%%%%%%

According to momentum and energy conservation, the QPM condition for PDC is given by
%%%%%%%%%%%%%%%%%%%%%%%%%%%%%%%%%%%%%%%%%%%%%%%%%%%
\begin{eqnarray}
\vv{ k}\left( {\omega _p=\omega _s+\omega _i} \right) = \vv{ k}\left( {\omega _s} \right) + \vv{ k}\left( {{\omega _i}} \right)+\vv{ k}_{G}\left( {m,\Lambda^{-1}} \right)
\label{eq:QPM}
\end{eqnarray}
%%%%%%%%%%%%%%%%%%%%%%%%%%%%%%%%%%%%%%%%%%%%%%%%%%%
where $\vv{k}(\omega)$ are the wave vectors of the involved    pump, signal and idler waves denoted by the subscripts p, s and i, respectively, $\vv{k}_{G}$ is the QPM grating vector with $|\vv{k}_{G}|=2\pi m/\Lambda$,  $m$ is the order of the interaction and $\Lambda$ is the poling period.
Via the nonlinear $\chi^{(2)}$ interaction photon pair states
%%%%%%%%%%%%%%%%%%%%%%%%%%%%%%%%%%%%%%%%%%%%%%%%%%%
\begin{eqnarray}
&&{\left| \Psi  \right\rangle _{PDC}} \propto \int\!\!\int d{\omega _s}d{\omega _i}{f \left( {{\omega _s}{\rm{,}}{\omega _i}} \right)} \hat a_s^\dag \left( {{\omega _s}} \right)\hat a_i^\dag \left( {{\omega _i}} \right)\left| 0 \right\rangle,
\label{eq:PDCstate}
\end{eqnarray}
%%%%%%%%%%%%%%%%%%%%%%%%%%%%%%%%%%%%%%%%%%%%%%%%%%%
are generated. Their spectral-temporal properties are determined by the joint spectral amplitude (JSA) ${f \left( {{\omega _s}{\rm{,}}{\omega _i}} \right)}=\alpha(\omega_s+\omega_i)\phi(\omega_s,\omega_i)$, where $\alpha(\omega_s+\omega_i)$ is the spectral pump amplitude distribution and $\phi(\omega_s,\omega_i){\rm{ = }} {\rm{sinc}} \left[ {\Delta{\it{k}} (\omega_s,\omega_i){\frac{\rm{L}}{2}}} \right]$ expresses the momentum conservation among the three fields in terms of the phase mismatch $\Delta k(\omega_s,\omega_i)$ .  The JSA contains full information about the bandwidth, the waveform and the type of frequency correlations of the two-photon state.

A basic requirement for many quantum protocols is generation of deccorelated photon pairs, sometimes also called separable, which means detection of one photon reveals no information about its partner. Spectral-temporal correlations can be characterised through JSA. In Fig. \Fig{QPM_JSI}~(a) detection of signal photon at a higher frequency tells us that the idler photon must be at a lower frequency, hence this is a correlated JSA; while Fig. \Fig{QPM_JSI}~(b) depicts a decorrelated JSA. It is a challenging task to generate separable PDC photons since the common non-linear materials result in a correlated JSA; thus, dispersion engineering has been a hot research topic for the past decade \cite{AnsariOp2018}. Even with most recent technological advancements, the generation of decorrelated photon pairs remains challenging and usually limited to specific wavelengths. As a result of this limitation, in many quantum optical experiments a tight spectral filtering is used, to achieve a decorrelated source, which in turn dramatically decreases the success rate of experiments.

Typically, co-propagating PDC is used to generate photon pair states, where all the waves propagate in the same direction. As shown in \Fig{QPM_JSI}~(a) the required grating compensation ${k}_{G}= {m2\pi}/{\Lambda}$ is relatively small, i.e., the poling period $\Lambda$ is correspondingly large. The phase mismatch is $\Delta k(\omega_s,\omega_i)=k_p(\omega_s+\omega_i) - k_s(\omega_s) -k_i (\omega_i)-k_G$. This can be seen from the JSA sketched in  \Fig{QPM_JSI}~(a), which looks like a tilted ellipse, e.g., typically PDC is not decorrelated caused by the dispersive properties of the material. Exceptionally, the generation of pure states with minimum correlations can only be achieved in very particular cases, e.g., for specific wavelength combination for which group velocity matching can be achieved \cite{KellerPRA1997}, and/or with strong spectral narrowband filtering, which results in a tremendous decrease of the effective efficiency of the source. Co-propagating decorrelated separable photon pairs \cite{HarderOE2013} can be generated via specific dispersion properties of given material, but with broad bandwidth and limited to a fixed telecom wavelength range. In fact, paradoxically, the spectral and the temporal properties of the photon pairs must be adapted to the desired application \cite{MunroNP2012,FicklerNC2014,IkutaNC2018}. Moreover, because of the relatively broad spectral width of the generated PDC photons, the temporal structure of the photon states can not be resolved as even the fastest single photon detectors have a limited temporal resolution in the range of tens of picosecond, which is still too slow for a time-resolved analysis. Thus, until now, energy-time entanglement for quantum information is only measurable using the Franson interferometry, but can not be directly observed due to the lack of resolution in time.

%%%%%%%%%%%%%%%%%%%%%%%%%%%%%%%%
\begin{figure*}[htb]
\center
\includegraphics[width=0.85\textwidth]{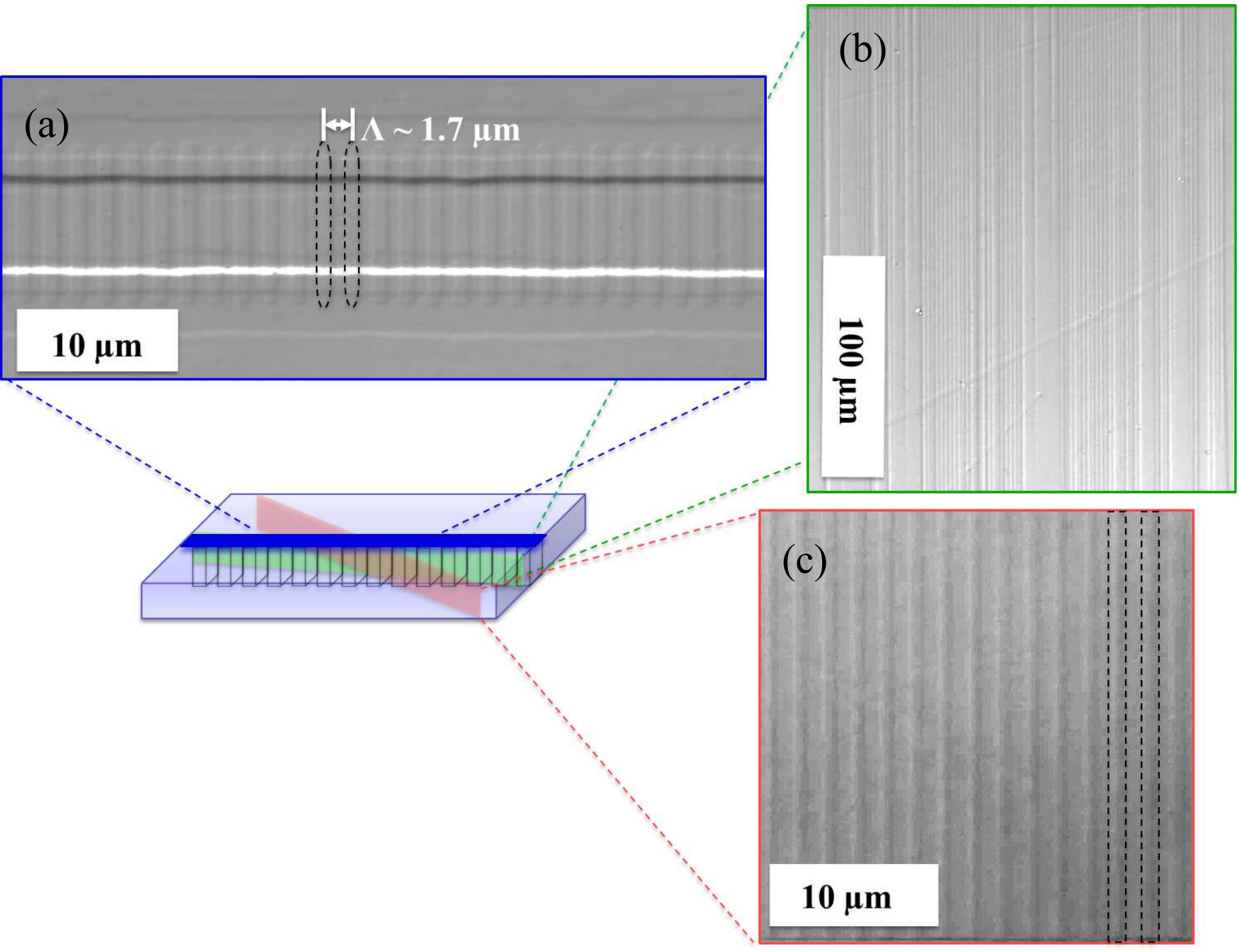}
\caption{\label{ShortPoling}
Micrographs of poled domain structures with 1.7 $\mu$m period. The measurement has been performed on a witness sample fabricated simultaneously with the sample that was used for the counter-propagating PDC generation. For visualization, we cut the sample along the sketched planes and subsequently etched the surface in hydrofluoric acid to reveal the domain structure exploiting selective etch rates for opposite domain orientations. The black dotted lines denote the inverted domains. (a) Top view onto the Ti:PPLN waveguide with a zoom to the domain structures. Oblique side views (b) an (c) with different polishing angles through the Ti:PPLN waveguide with a zoom showing the depth of poling structures.}
\end{figure*}
%%%%%%%%%%%%%%%%%%%%%%%%%%%%%%%%

All of these problems addressed above can be solved by exploiting a counter-propagation QPM configuration. The corresponding phase-matching requires a large $k_{CG}$ and, thus, an ultra-short poling period (\Fig{QPM_JSI}~(b)). For this counter-propagating scheme, the phase mismatch is $\Delta k_{C}(\omega_s,\omega_i)=k_p(\omega_s+\omega_i) - k_s(\omega_s) +k_i (\omega_i)-k_{CG}$, thus, the phase-matching function in the JSA diagram is almost horizontal and the spectral width of the idler photon is very narrow. More specifically, the spectral width is inversely proportional to the sum of the signal and idler group indices $|{{n_g}\left( {{\lambda _s}} \right) + {n_g}\left( {{\lambda _i}} \right)}|$, which is several orders of magnitude larger than in the co-propagating PDC case, where the difference  $| {{n_g}\left( {{\lambda _s}} \right) - {n_g}\left( {{\lambda _i}} \right)}|$ determines the spectral width.

Using a pulsed, i.e., broadband, pump the PDC process generates an almost decorrelated twin-photon state with a narrowband counter-propagating idler photon and a broadband co-propagating signal photon. For an intuitive understanding we must be aware that pump and signal propagate with almost the same speed through the waveguide. Thus, the generated signal temporarily overlaps with the pump pulse during the interaction process. The temporal width of the signal pulse is about the pump pulse length which can be much shorter than the propagation time through the waveguide. The idler, however, can be generated all along the interaction length and, hence, leaving the waveguide at any time between launching the pump pulse and twice the propagation time through the waveguide. Correspondingly, the resultant idler pulse must be much longer than the signal pulse which explains the asymmetry in the spectral bandwidth of signal and idler. Such direct generation of decorrelated photon pairs with different spectral bandwidths -- one is spectrally narrow enough for addressing stationary qubits in quantum memories at ultraviolet or visible wavelengths, while the other is broadband and capable to bridge the flying qubit at telecom wavelength ranges -- is an important step for the realization of quantum repeaters and hybrid quantum networks.

\section{Results}

First, we fabricate the waveguide with short poling periods to achieve counter-propagating phase matching. Field-assisted domain inversion is a mature technique for periodic poling of \LN\ waveguides. Typically, poling for co-propagating phase matched processes can routinely be obtained as periods of several micrometers and more are required. However, one main technical challenge to obtain counter-propagating PDC is the large $k_{CG}$, i.e., the short poling period $\Lambda_{CG}$, to obtain the phase-matching. First order phase-matching, $m=1$, requires a poling period of about 340~nm for degenerate PDC in the telecom range. However, the spatial resolution of the lithography pattern and sideways domain growth during poling limit the minimum poling period. To avoid such ultra-short periods, higher order phase-matching can be exploited, but with the drawback of a reduced efficiency of the nonlinear conversion. Thus, the poling technology had to be optimized to yield periods as short as possible. We were finally able to fabricate periodically poled Ti-indiffused \LN\ waveguides with periods down to about 1.7 $\mu$m, which provides fifth-order QPM for photon pair generation exploiting the highest nonlinear coefficient $d_{33}$. The top view and detailed side views of the domain structure of the waveguide are shown in \Fig{ShortPoling}.

%%%%%%%%%%%%%%%%%%%%%%%%%%%%%%%%
\begin{figure*}[htb]
\center
\includegraphics[width=0.85\textwidth]{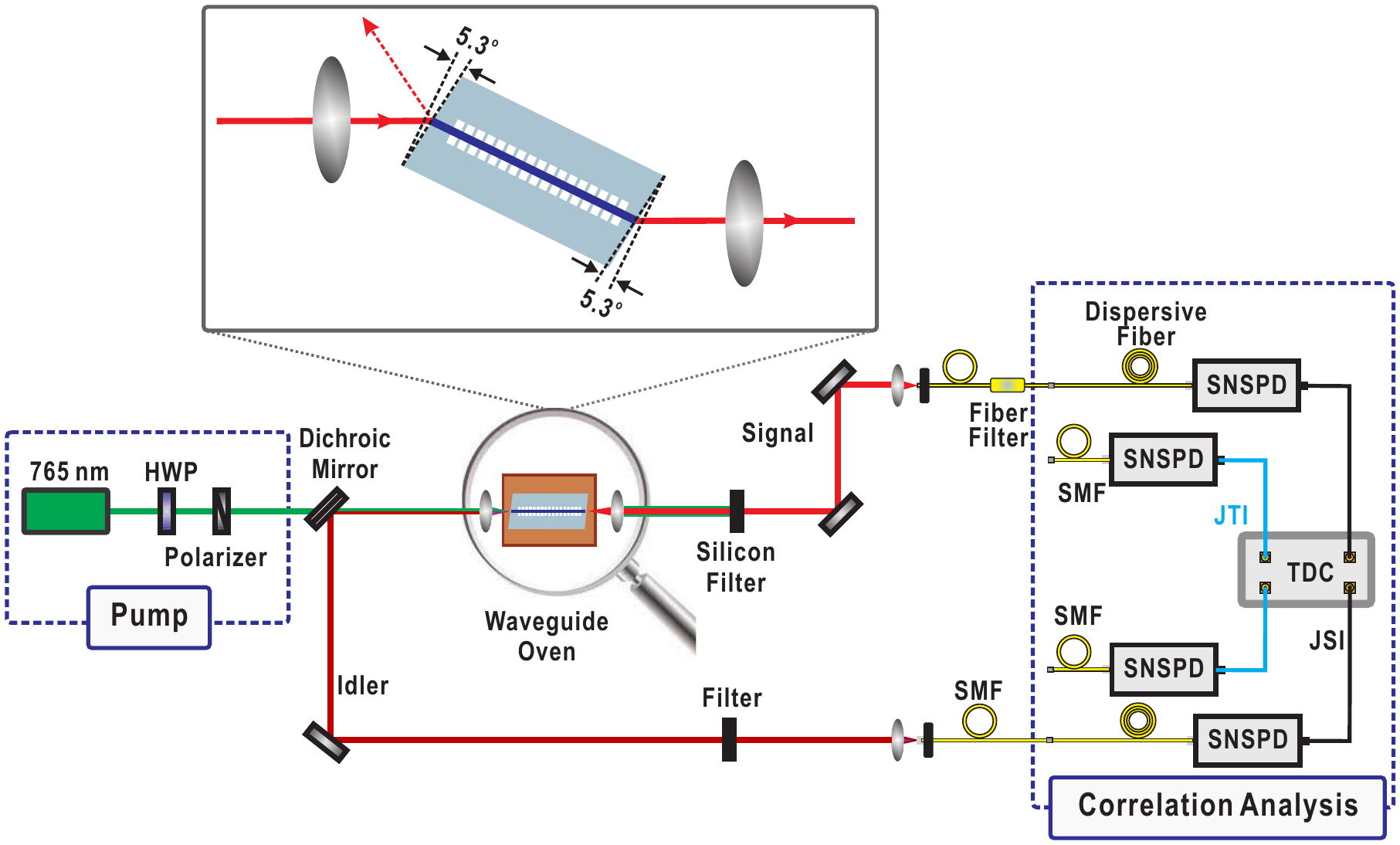}
\caption{\label{Setup}
Experimental setup for quantum characterization of the counter-propagating photon pair generation. The system is pumped with either a pump laser around 765 nm with pulse length of 2~ps and repetition rate of 80 MHz, or a continuous-wave laser. The end-face of waveguide are angle-polished. Pump power is set through a combination of half wave plate (HWP) together with a polarizer. In the inset, the coupling to the slant polished facets is visible. A temperature controller stabilizes the sample to  a temperature around 160~$^{\circ}$C to obtain quasi-phase-matching for the desired wavelength combination and to prevent luminescence and deterioration due to photo-refraction. During the measurements the sample temperature is stabilized to about $\pm$5~mK.
The two opposite output ports from the waveguide are coupled to single-mode fibres behind a filter stage consisting of a coated silicon filter to suppress residual pump light plus a 1.2 nm wide bandpass fiber filter to suppress background photons and a 8 nm wide bandpass filter in the signal and the idler arm, respectively. For the joint temporal intensity (JTI) measurements, the photons are detected with superconducting nanowire detectors (SNSPDs) and a time-to-digital converter (TDC). For the joint spectral intensity (JSI) measurements, long dispersive fibers are inserted in front of the detectors.
}
\end{figure*}
%%%%%%%%%%%%%%%%%%%%%%%%%%%%%%%%

Next, we investigated the quantum features of counter-propagating photon pairs which are generated inside the waveguide. To ascertain the twin character of the counter-propagating photon pairs, we performed both spectral and temporal correlation measurements. It is insufficient to characterize the generated state using coincidence measurements at the single photon level only, because a lot of processes could occur simultaneously due to inhomogeneity and imperfection of waveguide fabrication and periodic poling. For example, it is very difficult to certify that the main contribution of coincidence measurements from opposite directions after spectral filtering has excluded Cherenkov radiation \cite{BraschS2016}, multiple nonlinear conversion processes, and other back-reflection and noise (e.g., luminescence, photo-refraction, spontaneous scattering, etc.) which can occur inside waveguide simultaneously.

From the theoretical predictions, the spectral bandwidth of the counter-propagating idler photons should be very narrow, around the GHz range. Thus, it is very challenging to resolve the bandwidth with conventional spectroscopic methods at the single photon level. However, the spectral properties of the down-converted photons can be determined from temporal coincidence measurements, according to the Fourier relation between spectral bandwidth and temporal width. The narrower the photon bandwidth is, the broader the coincidence peak of the arrival time difference is. Besides, we also investigate the spectral and temporal correlations to certify the counter-propagating QPM.

%%%%%%%%%%%%%%%%%%%%%%%%%%%%%%%%
\begin{figure*}[htb]
\center
\includegraphics[width=0.85\textwidth]{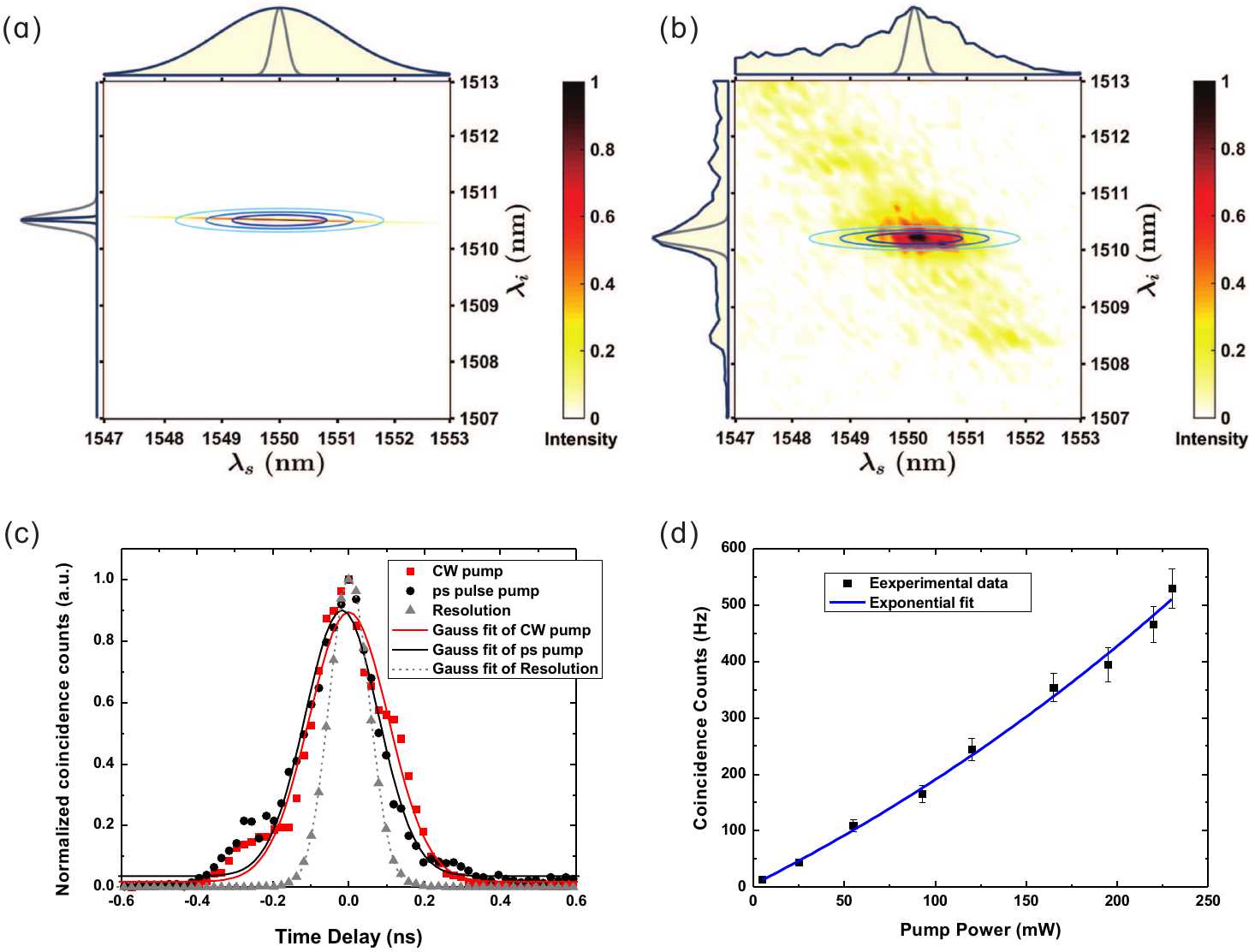}
\caption{\label{JSI} Pump independent heralded single photon source.
(a) Theoretically predicted counter-propagating joint spectral intensity (JSI) distribution between signal and idler. The theoretical JSI has asymmetric ellipse shape. The gray curve indicates the resolution of the detection system. The expected measured results defined by the detection resolution is shown by solid contour lines, from inside to outside representing the normalized intensity values 0.75 (navy), 0.5 (blue), and 0.25 (cyan), individually. (b) Experimental JSI measurements. The pump pulse duration is around 2 ps.
(c) Coincidence envelop es between signal and idler from ps pulse pumping (black) and continuous-wave (cw) pumping (red). The time resolution of the detection system (grey) was obtained by using a typical co-propagating QPM PDC source with short correlation time. (d) Dependence of coincidence counts on averaged pump power when pulse pump is used.
}
\end{figure*}
%%%%%%%%%%%%%%%%%%%%%%%%%%%%%%%%

The experimental set-up is shown in \Fig{Setup}. We studied counter-propagating PDC in a 37~mm long waveguide sample. It was pumped either by 2~ps long pulses from a Ti:sapphire laser or a cw-laser diode, both working at a wavelength of about 765~nm. To avoid spurious back reflections from the waveguide facets, waveguide endfaces were angle-polished at 5.3$^o$, which provides a suppression better than 30~dB. All coupling lenses have anti-reflection coatings for the telecom wavelength range. From the waveguide outputs to fiber coupling in front of each detector, the total transmission is around 50\% for signal and idler, which is including waveguide coupling, fiber coupling, and all optical components.

By using a fiber based time-of-flight spectrometer with long dispersive fibers, we first characterize the spectral properties of signal and idler, when it is pumped by the 2~ps pulsed laser. Since different wavelengths arrive at different times at super-conducting nanowire detectors (SNSPDs), through scanning the gating times of the time-to-digital converter (TDC), a joint spectral intensity (JSI) distribution is obtained from coincidence click rates as shown in \Fig{JSI}~(b). This is in good agreement with the theoretical expectation, as sketched in \Fig{JSI}~(a). Note the marginal spectral distributions of signal and idler are obtained from the single click rates. The spectral resolution of our fiber spectrometers is $\sim$ 0.2~nm, limited by the timing jitter of the SNSPDs and the length of the dispersive fiber.

In \Fig{JSI}~(c) we present the results of coincidence measurements in temporal domain for pulsed and cw pumping.
Additionally, the result from a co-propagating PDC source is shown for comparison. The latter source has a broad spectrum and, thus, a narrow coincidence width of a few ps only. From the observed width we determined the resolution of the detection system to be about 120~ps with an almost perfect Gaussian shape as indicated by the fit function. The measured width of the coincidence peak from the counter-propagating photons is about 250~ps, which fits well with expected width for the 37~mm long interaction length. This corresponds to a bandwidth of 1~GHz. The widths of the coincidence peaks for cw and ps-pumping are very similar. There is almost no broadening for pulsed pumping, since the width of the PDC photons is mainly determined by the horizontal phase-matching function, as shown in \Fig{QPM_JSI}. The slight shift of the peaks mainly comes from alignment issues when the two different pump systems are used.

According to the coincidence measurements, we also determined the efficiency of the counter-propagating PDC process. From the pump power dependent measurement as shown in \Fig{JSI}~(d), an intrinsic generated photon pair rate inside the waveguide of about 5 pairs/($\rm{s\cdot mW}$) is determined. We found that the pump power is still far below the threshold of a mirrorless OPO~\cite{CanaliasNP2007}, since the coincidence counts change with pump power almost linearly. Moreover, the non-quadratic behaviour reveals also that no other cascaded process contributes to the observed coincidences. Our brightness is about 5 orders of magnitude lower than the reported brightness using a co-directional PPLN waveguide source \cite{FujiiOE2007}.
This can be understood from the technological challenges for the realization of the sources. A higher order phase-matching has the drawback of a strongly reduced efficiency, which can be estimated from a Fourier analysis of the domain pattern along the interaction length. The efficiency scales with the square of the respective Fourier coefficient. For ideal rectangular poling patterns with 1:1 duty cycle, the efficiency drops proportional to $1/m^2$ with $m$ being an odd order.
Thus, a significantly enhanced efficiency can be expected if future progress in the poling technology enables low order phase-matching.
Furthermore, currently  non-ideally shaped poling patterns, which can arise if the domain flipping along the interaction length is not really sharp, further limit the achievable efficiency. In particular,
the transition between upwards- and downwards oriented domains has a tremendous impact on the Fourier components, which causes the decrease of generation efficiency.

%%%%%%%%%%%%%%%%%%%%%%%%%%%%%%%%
\begin{figure*}[htbp]
\center
\includegraphics[width=0.85\textwidth]{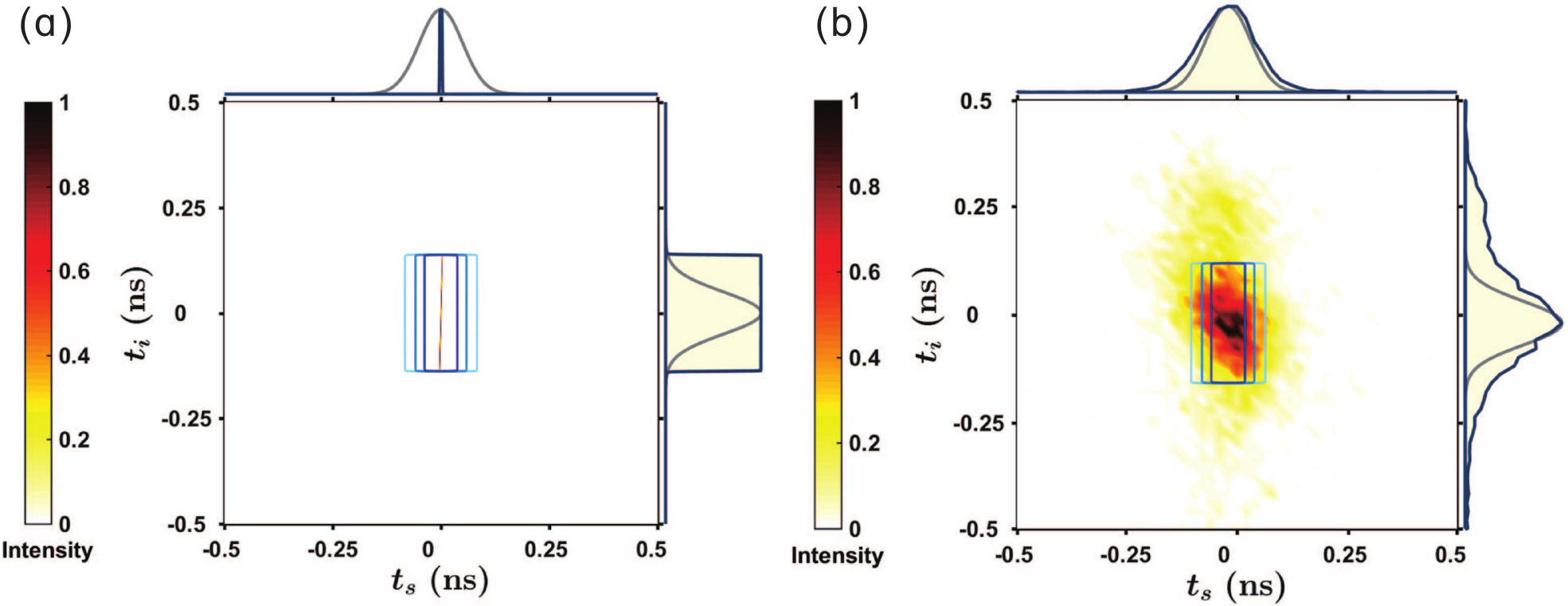}
\caption{\label{JTI}
Joint temporal intensity (JTI) results. (a) Theoretical prediction of counter-propagating JTI distribution between signal and idler. (b) Experimental result from JTI measurement, when pulse pump with duration 2 ps is used. The grey curve indicates the resolution of detection systems. The expected measured results defined by the detection system is indicated by solid contour lines, representing the normalized intensity values from inside to outside 0.75 (navy), 0.5 (blue), and 0.25 (cyan), respectively.
}
\end{figure*}
%%%%%%%%%%%%%%%%%%%%%%%%%%%%%%%

The spectrally narrow backward propagating idler photon with its longer life time \cite{MarianS2004} allows the direct measurement of the joint temporal intensity (JTI). This is completely different to conventional PDC cases for which such measurements  can only  be done either with  the help of pump delay \cite{KuzucuPRL2008} or up-converted detection \cite{MacLeanPRL2018}. The expected JTI has a rectangular shape according to the Fourier transform of the theoretical JSI. In principle, counter-propagating QPM in the nonlinear waveguide has decorrelated temporal intensity distributions, which is shown in \Fig{JTI}. The theoretical JSI has an elliptical shape with asymmetric axes (in \Fig{JSI}), while the JTI (as the Fourier transform of JSI) has rectangle profile with reversed asymmetric axes. When we take into account the resolution of detection systems (shown as grey curve in the figure), the expected measured results defined by the detection resolution is shown by solid contour lines. However, due to the temporal artifact of electric pulses from time converter systems, we cannot resolve the JTI shape clearly. That means, the directly measured JTI distribution cannot be resolved, because of the detector response function and the electronic noise of the TDC. Thus, we need to look at both the horizontally aligned elliptical JSI and the vertically aligned JTI very carefully to prove that the observed photon pairs are generated via the counter-propagating process and not a result of any spurious collinear processes.

Although the tendency of the main ellipse is parallel to the axis, there are multiple spectral structures along the horizontal direction in \Fig{JSI}~(b). From this JSI distribution and the asymmetrical tails of the coincidence peaks, the marginal spectral distribution of the signal and the marginal temporal distribution of the idler, we find that imperfections affect the observed results. This is mainly caused by inhomogeneities arising during waveguide fabrication and poling. For example, random QPM grating errors will impact the generation efficiency and cause more background. Furthermore, the waveguide inhomogeneities, like the waveguide width error due to the limitation of lithography resolution could also play a role.

\section{Discussions}

Our demonstration of counter-propagating QPM in a Ti:PPLN waveguide with the poling period on the order of the wavelength, and the generation towards narrow separable quantum states provide a new route for photonic quantum information applications. Our work shows that counter-propagating QPM in this long waveguide restricts the whole conversion spectrum towards pure single mode with GHz bandwidth of the counter-propagating photon and THz bandwidth of the co-propagating one. Our source exhibits prominent single mode behaviour in the frequency domain and remarkable temporal features, namely the temporal bandwidth is beyond the timing jitter of detector systems. In particular, we observe both the spectral decorrelation and temporal broadening of the counter-propagating PDC state. Therefore, it will benefit optical quantum information communication and processing which rely on pure and condensed generation of single photons with a coherence time longer than the time jitter of detector. Moreover, counter-propagating PDC is based on single pass geometry. This features several distinct advantages, which is completely different with narrowband photon pair generation via cavity-enhanced co-propagating PDC. Waveguide losses are less critical due to the single-pass operation and requirements on temperature stability are stringent compared to the multi-pass cavity sources. This means that, as soon as we have achieved the sub-$\mu$m poling technology, the interface between flying qubits and stationary qubits in different wavelengths and corresponding bandwidths should be available and easily operable.

Counter-propagating QPM achieved by using short poling periods in waveguides has great potential for various applications. For instance, it becomes the most promising way to interface various quantum memories with fiber networks, because of heralded decorrelated states, excellent temporal behavior and good compatibility with fiber.

We have demonstrated that it is possible to produce short poling periods in lithium niobate waveguides and showed the generation of counter-propagating photon-pairs with their specific properties.  With further advances in the poling technology, we expect this this will become a valuable tool for the engineering of quantum states. We anticipate that counter-propagating QPM will become a standard means in future quantum optics experiments, and will be used extensively in nano-photonic quantum information systems as well. These ideas can be extended to implement counter-propagating entangled state generation \cite{Liu2019} by using lithium niobate thin films combining with high-speed nano-photonic electro-optic modulators \cite{WangOE2018}.

\section*{Funding}
Deutsche Forschungsgemeinschaft (DFG -- German Research Foundation) (231447078 -- TRR 142 C02, B05); Gottfried Wilhelm Leibniz-Preis (SI1115/3--1).

\section*{Acknowledgments}
We thank B. Brecht and V. Quiring for helpful discussions and technical assistance, and E. Meyer-Scott and M. Stefszky for proofreading the manuscript.

\section*{Disclosures}

\textbf{Author contributions:} K.-H.L., V.A. and M.M. built the experiment and collected the data. C.E. and R.R. contributed to the fabrication of sample. K.-H.L, V.A., M.S. and H.H. provided theoretical support. K.-H.L., H.H. and C.S. conceived the original idea. C.S. supervised the project. K.-H.L. and H.H. drafted the manuscript. All authors discussed and contributed to the final version of the manuscript. \textbf{Competing interests:} The authors declare that they have no competing financial interests.

%%%%%%%%%%%%%%%%%%%%%%% References %%%%%%%%%%%%%%%%%%%%%%%%%

\bibliographystyle{}

\end{document}